\documentstyle[aps,prl,epsfig]{revtex}

\begin{document}

\newcommand{\be}{\begin{eqnarray}}
\newcommand{\ee}{\end{eqnarray}}
\newcommand{\beq}{\begin{equation}}
\newcommand{\eeq}{\end{equation}}
\newcommand{\xx}{\begin{eqnarray*}}
\newcommand{\yy}{\end{eqnarray*}}
\newcommand{\nn}{\nonumber}
\newcommand{\Vol}{{\rm Vol}}
\newcommand{\sign}{{\rm sign}}
\newcommand{\tr}{{\rm Tr}}

\twocolumn[\hsize\textwidth\columnwidth\hsize\csname@twocolumnfalse\endcsname

\title{ Universal Fluctuations in Correlated Systems }

\author{ S.T. Bramwell$^1$,K. Christensen{$^2$}, J.-Y. Fortin$^3$, P.C.W.
Holdsworth$^4$,
H.J. Jensen$^5$, S. Lise{$^5$}, J. L\'opez{$^5$}, M. Nicodemi{$^5$}, J.-F.
Pinton$^4$, M. Sellitto$^4$}

\address{$^1$ Department of Chemistry, University College London, 20
Gordon
Street, London, WC1H~0AJ, United Kingdom. \\
{$^2$}Blackett Laboratory, Imperial College,
Prince Consort Road, London SW7 2BZ United Kingdom.\\
$^3$ Department of Physics, University of Washington, Box
351560,Seattle, WA 98195-1560, USA. \\
$^4$ Laboratoire de Physique, Ecole Normale Sup\'erieure,
46 All\'ee d'Italie, F-69364 Lyon cedex 07, France.\\
$^5$ Department of Mathematics, Imperial College London, London, SW7,
United Kingdom.}

\maketitle

\begin{abstract}
The probability density function (PDF) of a global measure in a large
class of highly correlated systems has been suggested to be of the same
functional form. Here,  we identify the analytical form of the PDF 
of one such measure, the order parameter in the low temperature 
phase of the 2D-XY model. We demonstrate that this function describes 
the fluctuations of global quantities in other correlated, equilibrium 
and  non-equilibrium systems. These include a coupled rotor model, 
Ising and percolation models, models of forest fires, sand-piles,
avalanches and granular media in a self organized critical state. 
We discuss the relationship with both Gaussian and extremal statistics.

\medskip

\medskip

\noindent{PACS numbers: 05.40, 05.65, 47.27, 68.35.Rh}

\end{abstract}

\twocolumn\vskip.5pc]\narrowtext

\narrowtext

Self similarity is an important feature of the natural world. It
arises in
strongly correlated many body systems when fluctuations over
all scales from a microscopic length $a$ to a diverging correlation
length
$\xi$ lead to the appearence of ``anomalous
dimension''~\cite{Gold} and fractal properties.
However,
even in an ideal world the
divergence of of $\xi$ must ultimately be cut off by a macroscopic
length $L$,
allowing the definition of a range of scales between
$a$ and $L$, over which the anomalous behaviour can occur. Such
systems are found, for
example,  in critical phenomena, in Self-Organized
Criticality~\cite{Pbak,jensen1} or in turbulent flow problems. By analogy with
fluid mechanics we shall call these finite size critical systems
``inertial systems''
and the range of scales between $a$ and $L$ the ``inertial range''.
One of the anomalous statistical
properties of inertial systems is that, whatever their size,
they can
never be divided into mesoscopic regions that are statistically
independent. As
a result they do not satisfy the basic criterion of the central limit
theorem
and one should not necessarily expect global, or spatially averaged
quantities
to have Gaussian fluctuations about the mean value.
In Ref.~\cite{BHP}(BHP) it was demonstrated
that two of these systems, a model of finite size critical behaviour
and a steady state in a closed turbulent flow experiment,
share the same non-Gaussian
probability distribution function (PDF) for fluctuations of global
quantities.
Consequently it was proposed that these two systems - so utterly
dissimilar in regards to their microscopic details - share the same
statistics
simply because they are critical.
If this is the case, one should then be able to describe turbulence as
a finite-size critical phenomenon, with an effective ``universality
class''. As,
however, turbulence and the magnetic model are very unlikely to share
the same
universality class, it was implied that the differences that separate
critical
phenomena into universality classes represent at most a minor
perturbation on
the functional form of the PDF. In this paper, to test this
proposition, we
determine the functional form of the BHP fluctuation spectrum and
show that it
indeed applies to a large class of inertial systems \cite{Harte}.

The magnetic model studied by BHP, the spin wave limit to the two
dimensional XY (2D-XY) model, is defined by the harmonic Hamiltonian

\be \label{eq1}
H=-J\sum_{\langle i,j \rangle}\left[1- {1\over{2}}\left
(\theta_i-\theta_j\right )^2\right]
\ee

\noindent
where $J$ is the near neighbour exchange constant for angular
variables
$\theta_i$ that occupy a square lattice with periodic boundary
conditions.
The magnetization is defined as
$m = 1/N \sum_i \cos(\theta_i -\overline{\theta})$, where
$\overline{\theta}$ is the instantaneous mean orientation.
This model is critical at all temperatures and for an infinite
system has algebraic correlations on all length scales. In the finite
system the
lattice constant $a$ and the system sizes $L = a \sqrt{N}$ define a
natural inertial range. The model can be
diagonalized in Fourier space, which makes it very convenient for
analytical work.  The PDF of the magnetization $P(m)$ can be
expressed as
the Fourier
transform of a sum over its moments.
In Ref.~\cite{SF} it was shown that the moments are given by
$\mu_n =
g_n(g_2/2)^{-n/2}\sigma^n$, where $\sigma^2$ is the variance and the
$g_k (k = 2,
3, 4...)$ are sums related to the lattice Green
function in Fourier space $G({\bf q})$: $g_k=\sum_{{\bf q}}G({\bf
q})^k/N^k$.  The fact that $\mu_n \propto \mu_1^n$ means that a
change of
$N$ or
$T$ is equivalent to a linear transformation of the variate $m$;
hence, the
PDF can be
expressed in a universal form. As shown in \cite{SF} the
moment
series can
be resummed to give the following expression, exact to leading order
in $N$:

\be\label{eq2}
P(y)=\int_{-\infty}^{+\infty}\frac{dx}{2\pi\sigma}
\exp\left [iy x
+\sum_{k=2}^{\infty}\frac{g_k}{2k}
\left (ix\sqrt{\frac{2}{g_2}}\right )^k \right ].
\ee

\noindent
Here $y = (m- \langle m \rangle)/\sigma$ and $ \langle m \rangle $ is
the
mean of the
distribution.
Including only $g_2$ in (\ref{eq2}) would give a Gaussian PDF with
variance $\sigma^2$. However the terms for $k>2$
%are not of higher order $N$ and $T$; they
cannot be neglected and
$\Pi(y) = \sigma P(y)$ is a non-Gaussian, universal function,
independent of both the size of the system and the temperature.
Without loss of generality one can  make the quadratic approximation
$m=1-\sum_i (\theta_i-
\overline{\theta})^2/2N$
which allows us to
transform Eqn.(\ref{eq2}) to  a form suitable for numerical
integration~\cite{Fortin}:

\be\label{eq2b}
\Pi(y)= \int_{-\infty}^{+\infty}\sqrt{\frac{g_2}{2}}\frac{dx}{2\pi}
\exp\left [i \Phi(x)\right ]
\ee

\be
\nn
i \Phi(x) = ixy\sqrt{\frac{g_2}{2}}
-i\frac{x}{2}\tr G/N-\frac{1}{2}\tr\log\left ({\bf 1}-ixG/N\right)
\ee

\noindent
(Here the trace Tr of any function of $G$
is defined as the
sum for
${\bf q}\neq {\bf 0}$ of the same function of $G({\bf q})$, which can
be simply proved by making a Taylor expansion of the function near
$G=0$.)
In order to make an accurate test of this expression we have
performed a high
resolution molecular dynamics simulation of $P(m)$.
Fig.~1 compares the integrated equation \ref{eq2} with data for
a system of $1024$ classical rotors
integrated over
$10^8$
molecular dynamics time steps in the low temperature phase. The agreement is
globally excellent,
particularly in the wings of the distribution and along the exponential tail
for
fluctuations below the mean.

The asymptotic values of $\Pi(y)$ are related to the saddle points of
the integrand in (\ref{eq2b}). We find

\be\label{eq3}
\Pi(y)\propto |y|\exp (- \frac{\pi}{2} b y); \; {\rm for}\;y  \ll 1
\ee

\be\label{eq4}
\Pi(y)\propto \exp (
-\frac{\pi}{2}e^{b (y-s)}); \;{\rm for}\;y \gg 1
\ee

\noindent
where  $b = \frac{1}{8\pi} \sqrt{g_2/2} \approx 1.105$ and  $s =
0.745$.
These forms give the correct asymptotic gradients of the molecular
dynamics
data on logarithmic and double logarithmic scales. The asymptotic forms are
an accurate approximation to Eqn.
(\ref{eq2}) for large $|y|$; however deviations from the asymptotes
are important over most of the physical range of $y$, which is
typically  limited to $\log{|y|} \sim O(1)$.  Equations (\ref{eq3})
and
(\ref{eq4}) serve as a guide to finding a good approximation to the
functional form of $\Pi(y)$ in this range.  To do this, we observe
that the factor of $y$ in (\ref{eq3}) can be regarded as constant in
this regime, which along with (\ref{eq3}) and (\ref{eq4}) immediately
suggests the form

\begin{equation}
\label{BHP-eq}
\Pi(y) = K \left(e^{x -e^x}\right)^{a} ; x = b(y - s), \; a=\pi/2.
\end{equation}

\noindent
This function must obey the three conditions of unit area, zero mean
and
unit variance, which fixes $b$, $s$ and $K$ to values slightly
different to
those found analytically: $b = 0.938$, $s = 0.374$, $K = 2.14$. An
alternative approach is to choose the parameters in the generalized
function $N e^{a(b(y-s) -e^{b(y-s)})}$ such that the first four
Fourier coefficients match Eqn.(\ref{eq2}). In this case we find $ a =
1.58$, $K = 2.16$, $b = 0.934$, $s = 0.373$,  in satisfying agreement
with
the previous estimates. The ratios of the higher order Fourier
coefficients differ from unity only very slowly, showing
that Eqn.~(\ref{BHP-eq}) is an accurate approximation to $\Pi(y)$.
This is directly confirmed by plotting Eqn.~(\ref{BHP-eq}) versus the
molecular dynamics and exact results in Fig.~1, where the fit is
seen to
be of extremely high precision.

\begin{figure}[h]
\centerline{\epsfig{file=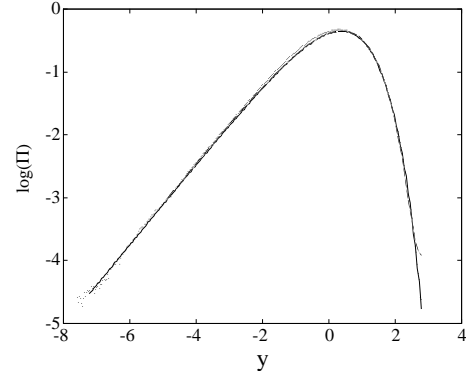,height=5cm}}
\caption{{\small The PDF $\Pi (y) $  as found from a numerical Fourier transform
of Eqn.~(\ref{eq2b}) (long dash), from Eqn.~(\ref{BHP-eq}) (solid)
and by Molecular Dynamics simulation of a system of $N=1024$ classical
rotors (dotted). }}
\end{figure}

We now test the idea that the BHP fluctuation spectrum of the form of
Eqn.~\ref{BHP-eq} is exhibited by many types of inertial system.
Fig.~2 shows the numerically simulated PDF of global quantities in several
equilibrium and non-equilibrium models. The equilibrium models include the
2D Ising model at a temperature $T^{\ast}(N)$ 
just below the critical temperature and a 2D site percolation model on a
square lattice for a site occupation probability $P^{\ast}(N)$ just above the
percolation transition. The numerical results refer to the fluctuations of
the absolute value of the magnetization and the fluctuations in the size of
the spanning cluster respectively. The non-equilibrium models are of
the type that when driven slowly enter a scale free or critical
steady state defined as
Self-Organized
Criticality (SOC)~\cite{Pbak,jensen1}. Here the global quantity
is essentially a dissipation
rate that fluctuates about a well defined mean value in the steady
state. Details of the
individual SOC models
are as follows. (1) {\em The auto igniting forest fire
model}~\cite{proshun}
consists of ``trees'' planted at random on the vacant sites of a
square lattice  with
probability  $p$. In each time step the age $T_i$  of a tree on site
$i$ is
incremented by one unit. When $T_i = T_{max}$ the
tree ignites and $T_i$ is reset to zero.
Trees can also catch fire by being nearest neighbour to a site on
fire. The energy, or
wood stored in a tree, is proportional to $T$, and the figure shows the PDF
of the total energy dissipated in fires at each time step.
(2) In the {\em Bak-Tang-Wiesenfeld (BTW) sandpile
model}\cite{BTW} a dynamical variable $E_i$ is defined on
lattice site $i$.
The model is driven by adding units of the $E$-field to randomly
selected
sites.  When $E_i > E_{max}$ the site variable is decreased by
$E_{max}$
and the $E$-variable of the $z$ neighbour sites are increased by
$E_{max}/z$. One
or more of the neighbour sites may then aquire an $E$-value larger
than
$E_{max}$ and an avalanche is induced.  The PDF shown refers to the
fluctuations
in the instantaneous number of relaxing sites.
(3) In the {\em Sneppen depinning model}\cite{sneppen} an interface
moves through
a static random  field of pinning forces. The site along the
interface that experiences the smallest pinning force is moved one
unit ahead. If the local slope $s_i$ exceeds $1$, then the
neighbouring sites are moved one unit
ahead until all $s_i \le 1$.
We call such a sequence of updates a micro-avalanche and calculate the
PDF of the sum of areas covered by the progressing interface during an
integral time scale $T$ dependent on system size.
(4) The model for {\em granular media} is a `Tetris-like'  2D lattice gas
ensemble of anisotropic particles settling under gravity in a finite box
~\cite{granular}.
Due to the geometrical frustration, the total mass varies from one
realization of the filling process to another. The PDF for
fluctuations in bulk density of the particles is shown.

\begin{figure}[h]
\centerline{\epsfig{file=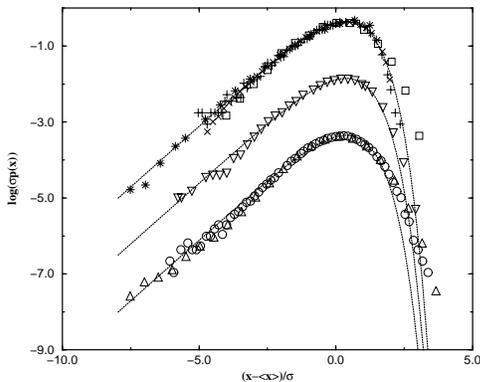,height=5cm}}
\caption{{\small Fluctuation spectra in equilibrium systems (lower 
curves): Ising ($\circ$), percolation ($\triangle$). 
The central curve ($\nabla$) correspond to the fluctuation spectrum 
of the correlated extremal process -- see 
text. The upper curves are the PDFs for the auto igniting forest fire 
model ($\times$), the Sneppen depinning model ($+$), the granular 
media model ($\Box$) and the BTW sandpile model ($*$).  The dotted line
corresponds to Eqn.~(\ref{BHP-eq}). For clarity, the sets of curves 
are shifted downwards by 1.5 in log units. }}
\end{figure}

Refering to Fig.~2, the
data sets for all models fall close to the BHP form, Eqn.~(\ref{BHP-eq}). In
the equilibrium models (lower curves) the self similarity is expected at 
the  system-size
dependent critical temperature $T^{\ast}(N)$ or percolation 
probability $P^{\ast}(N)$ only. The PDF for
 the 2D Ising model, for example, is temperature dependent, but makes a
close approach to the BHP form around $T^{\ast}(N)$.
We believe that the remaining deviations for fluctuations above the mean are
due to the limited inertial range for the system sizes studied. 
For the non-equilibrium systems (upper curves) some of the data sets
also show some deviation.  This may simply 
be due to poor statistics, as the deviations on the two sides of the
mean are related by the constraints of normalisation. In this respect, we
note that extremely good
statistics were required to get a really satisfactory fit to
Eqn.~(\ref{BHP-eq}) for the 2D-XY model,
while for more limited data sets systematic deviations occurred. 

Nevertheless it is clear that, to leading order, Eqn.~(\ref{BHP-eq}) 
correctly gives the behaviour of the global fluctuation spectrum in all 
these systems,
independently of the details of the each example. We propose that this is a
consequence of the systems sharing the properties of finite size, strong
correlations and self similarity.

To clarify this proposition, we return to our
calculation on the 2D-XY model. One can see explicitly that the BHP
spectrum occurs through appearance of anomalous dimension and contributions
on all
length scales of the intertial range.
The magnetization can be written, within the quadratic approximation,
$m = 1 -  \Sigma_{\bf q} m_{\bf q}$, where the $m_{\bf q}$s are the
amplitudes from the individual spin wave modes. These are
statistically independent positive variates with PDF
\beq\label{micro}
  P(m_q) = \sqrt{\beta q^2 N\over{4 \pi}} m_q^{-1/2} \exp(-\beta Nq^2 m_q),
\eeq
whose mean and standard deviation therefore scale with $q^{-2}$. The
``softest''
modes have
wave vector $q=2\pi/L$ and hence, by themselves, make contributions
of  $O(1)$
to $m$, while the modes on the zone boundary with $q = \pi/a$ have
only
microscopic amplitude. The moments of $P(m)$ are determined by
the mean magnetization. This is proportional to the integral over all
contributions: $\sim \int_{2\pi/L}^{\pi/a} q^{-2}n(q) dq$ where $n(q)
\sim
q^{d-1}$ is the density of states.
In one dimension the integral depends only on the lower limit
$2\pi/L$ and
only the soft modes count, while in three dimensions only the upper
limit
$\pi/a$ is
important and the multitude of modes near the zone boundary dominate
the
sum. In
two-dimensions however, both
limits of the integral are required and a detailed calculation gives
$\langle m \rangle  = 1 -{\eta/2}
\log{(CL/a)}$, with $C=1.87$~\cite{Fortin} and critical exponent
$\eta = T/2\pi J$.
The relevance of fluctuations over all length scales
of the zone therefore leads to the ``anomalous'' term $\log{(L/a)}$
and it ensures that the system cannot be cut into statistically independent
parts.

The spinwave approximation to the XY-model is exactly equivalent to the
Edwards-Wilkinson (EW)
model of a growing interface in steady state~\cite{EW}, with the square of
the interface width
$w$ equal to the sum over the amplitudes $m_q$: $w^2 = \sum_q m_q$. The
fluctuations in the
width of the interface has been studied by Foltin et al.\cite{foltin} for
the 1D case and by
R{\'a}cz and Plischke~\cite{racz} for the 2D interface.
The BHP spectrum is found for the {\it critical} two dimensional 
case only and our
calculation can be considered the completion of the study in Ref.~\cite{racz}.

The functional form of Eqn.~(\ref{BHP-eq}) suggests a
relationship to
Gumbel's first asymptote~\cite{Gumbel} for extreme value statistics, which
have recently been discussed in relation to turbulence in one
dimension\cite{bouchaud}. The form (\ref{BHP-eq}) but with $a$ taking
{\em
integer} values, where
$a = 1, 2, 3...$ would correspond to the PDF for the first, second,
third.... largest of
the $N$ random numbers. However, the exponent $a=\pi/2$ suggests, as
we have
argued, that the fluctuations in $m$ are not dominated by single
independent variables.
Rather, the analytic derivation of Eqn.~(\ref{BHP-eq}) shows that if
extreme
value statistics are involved they must be related to the
statistics of some emergent coherent collective excitation of the
system.
This is born out in the simulations of the Ising model and of all the SOC
models studied. In the Ising model, it is found that both the full
magnetization and
the contribution to the magnetisation from the largest
connected cluster of parallel spins give the same PDF, within numerical
error. For the Sneppen model the PDF of the {\it sum} over avalanches and
that of the {\it
largest} avalanche during the
integral time $T$, are both found to be of the BHP form,
even though these quantities are not related by a simple scale.
If the avalanches appearing during time
$T$ were uncorrelated
one would expect that the PDF for the largest avalanche would be
Gumbel's asymptote with $a=1$. The modification towards our form
indicates therefore that there are correlations between events during
the period $T$.
To test this idea, we have studied the PDF of the extreme values
taken from
sets of linearly
correlated variables.
The  process consists of generating a vector
$\vec{\chi}=(\chi_1,...\chi_N)$ where
$\chi_i$ are all independent and exponentially distributed. The
maximum signal
is obtained as
$\xi_{max} = \max\{\xi_1,...,\xi_N\}$ where the vector $\vec{\xi}={\bf
M}\vec{\chi}$
and ${\bf M}$ is an $N\times N$ matrix with random but fixed
elements. The resulting spectrum, the central data set in Fig.~2,  
is found to be very close to the BHP spectrum.

In conclusion, our results
infer that the non-Gaussian PDF of a global quantity in a critical system
is a consequence of finite size, strong correlations and self similarity, and
is independent of universality class to leading order.
Clearly many more studies of this point are required.
Non-linearity does not appear to be an essential feature, over and
above the necessity, in a closed system, to couple the elementary degrees of
freedom.
Indeed,  if non-linearity were essential, it would seem
impossible that the linear spin-wave theory could capture the fluctuations
in the
turbulence experiment \cite{BHP}.
R{\'acz} and Plischke~\cite{racz} have studied a series of linear and
non-linear models
for growing interfaces.
All show anisotropic PDFs for the interface width, with long tails in
qualitative agreement
with our data.
It would very interesting to examine these models in detail to see how,
in a controlled
environment the strength of the non-linearity affects the form of the
fluctuations.
Finally, it seems that a relationship exists between the BHP curve and
extremal statistics.
Although we have shown that the BHP behavior is not simply due to extreme
values of the statistically
independent degrees of freedom of the 2D-XY model, extreme values do appear
to dominate the
real space coherent structures that are excited in the critical (Ising
model) or
self-organized critical (Sneppen model) state.

Our findings thus establish a completely new and general consequence of
self similarity and they open the door to numerous studies that could lead
to a unified global description of aspects of 
equilibrium and non-equilibrium behaviour.

It is a pleasure to thank P. Sinha-Ray for supplying the auto
igniting forest fire model data. This work was supported by the
Minist\`ere de l'Education et de la Recherche No. ACI-2226(2), 
HJJ and SL are supported by the British EPSRC, JL and MS
(ERBFMBICT983561) are supported by the European Commission.

\end{document}